\documentclass[final]{svjour3}
\usepackage{graphicx}
\usepackage{rotating}
\usepackage{amssymb}
\usepackage{mathptmx}
\usepackage[numbers]{natbib}
\makeatletter
\journalname{Journal of Low Temperature Physics}

\bibpunct{}{}{,}{s}{}{,}
\usepackage{url}

\begin{document}

\newcommand{\hdblarrow}{H\makebox[0.9ex][l]{$\downdownarrows$}-}

\title{The Cryogenic AntiCoincidence detector for ATHENA X-IFU: assessing the role of the athermal phonons collectors in the AC-S8 prototype}

\author{
M.~D'Andrea\textsuperscript{1,2} \and C.~Macculi\textsuperscript{1} \and A.~Argan\textsuperscript{1} \and S.~Lotti\textsuperscript{1} \and G.~Minervini\textsuperscript{1} \and L.~Piro\textsuperscript{1} \and M.~Biasotti\textsuperscript{3} \and D.~Corsini\textsuperscript{3} \and F.~Gatti\textsuperscript{3} \and G.~Torrioli\textsuperscript{4}
}

\institute{ 
\email{matteo.dandrea@iaps.inaf.it}\\
\textsuperscript{1}INAF/IAPS Roma, Rome, Italy\\
\textsuperscript{2}Dept. of Physics, University of Roma "Tor Vergata'', Rome, Italy\\
\textsuperscript{3}Dept. of Physics, University of Genova, Genoa, Italy\\
\textsuperscript{4}CNR/IFN Roma, Rome, Italy
}

\titlerunning{The CryoAC for ATHENA X-IFU: role of the athermal phonons collectors in AC-S8}      
\authorrunning{M. D'Andrea et al.}

\maketitle

\begin{abstract}
The ATHENA X-ray Observatory is the second large-class mission in the ESA Cosmic Vision 2015-2025 science programme. One of the two on-board instruments is the X-IFU, an imaging spectrometer based on a large array of TES microcalorimeters. To reduce the particle-induced background, the spectrometer works in combination with a Cryogenic Anticoincidence detector (CryoAC), placed less than 1 mm below the TES array. The last CryoAC single-pixel prototypes, namely AC-S7 and AC-S8, are based on large area (1 cm$^2$) Silicon absorbers sensed by 65 parallel-connected iridium TES. This design has been adopted to improve the response generated by the athermal phonons, which will be used as fast anticoincidence flag. The latter sample is  featured also with a network of Aluminum fingers directly connected to the TES, designed to further improve the athermals collection efficiency. \newline
In this paper we will report the main results obtained with AC-S8, showing that the additional fingers network is able to increase the energy collected from the athermal part of the pulses (from the 6\% of AC-S7 up to the 26 \% with AC-S8). Furthermore, the finger design is able to prevent the quasiparticle recombination in the aluminum, assuring a fast pulse rising front (L/R limited). In our road map, the AC-S8 prototype is the last step before the development of the CryoAC Demonstration Model (DM), which will be the detector able to demonstrate the critical technologies expected in the CryoAC development programme.


\end{abstract}

\section{Introduction}

The Advanced Telescope for High ENergy Astrophysics (ATHENA\cite{athena}) is the second large-class mission selected in ESA Cosmic Vision 2015-2025. It will operate at L2, the second Lagrange point of the Sun-Earth system, with the launch planned in 2028. The mission is designed to address the scientific theme ``The Hot and Energetic Universe''\cite{heuniverse}, performing X-ray observations in the range 0.2-12 keV. The  X-ray Integral Field Unit (X-IFU\cite{XIFU}) is one of the two instruments of the payload. It is a cryogenic spectrometer that will provide spatially resolved high-resolution spectroscopy ($\Delta$E $<$ 2.5 eV at 6 keV) over a 5 arcmin diameter Field Of View. The core of the instrument is a large array (about 4000 pixels) of Transition Edge Sensor (TES) microcalorimeters, operated with a thermal bath around 50 mK.

The X-IFU performances are strongly influenced by the  particle background expected in the L2 environment, which is induced by minimum ionizing particles such as protons (of both galactic and solar origin) and secondary electrons\cite{back}. Background reduction techniques have been adopted to reduce this contribution by a factor $\sim$100 down to the requirement of 0.005 cts/cm$^2$/s/keV. These techniques mainly consist in the adoption of a passive electron shielding surrounding the TES-array, and a TES-based Cryogenic AntiCoincidence detector (CryoAC). Detailed Monte-Carlo simulations drive the CryoAC development, identifying the requirements that translate into the detector specifications\cite{CryoACspie}. The baseline CryoAC design foresees 4 identical pixels placed less than 1 mm below the TES array absorbers. Each pixel has an active area of $\sim$ 1.2 cm$^2$, a dynamic energy range from 5 keV to 750 keV (TBC) and a separated readout chain based on a DC-SQUID.

Our consortium has developed and tested several CryoAC prototypes so far, achieving a TRL (Technology Readiness Level) $\sim$ 4 at the single-pixel level\cite{LTD16}$^,$\cite{LTD15}. The next milestone in our schedule is the development of the CryoAC DM. It will be a single pixel (1 cm$^2$ area) aimed to probe the detector critical technologies, in particular the operation with a 50 mK thermal bath and the low energy threshold at 20 keV. In order to reach the final DM design we have developed two ``pre-DM'' samples, namely AC-S7 and AC-S8. The characterization and test activities performed on AC-S7 have been already reported in \cite{PCA}$^,$\cite{spiebia}. Here we present the main results obtained with the latter sample: AC-S8.

\section{The last-generation CryoAC prototypes: AC-S7 and AC-S8}

The two ``pre-DM'' prototypes (Fig.~1a,b) have been produced at the Genova University (Phys. Dept.). Their structure is based on a large area (1 cm$^2$) Silicon absorber sensed by a network of 65 Ir TES in parallel connected through Nb lines, and readout as a single object. To ensure a reproducible thermal conductance towards the thermal bath, the absorber is connected to a Silicon buffer (in strong thermal contact with the bath) via 4 SU-8 hollow towers filled with epoxy glue (Fig.~1c). The shape of the towers allows us to control their thermal conductance, which is expected to be some 10$^{-8}$ W/K @ 100 mK. The main parameters of the samples are reported in Tab.~1. 

The energy deposited in the absorber by particles and photons is released in form of electron-holes pairs and phonons (about 70$\%$)\cite{moseley}. Part of this energy (a few percent) goes on high energy phonons that travels in quasi-ballistic way (Fig.~1d). These phonons can enter the TES, being efficiently absorbed by the free electrons of the metal film\cite{probst}. This interaction quickly heats the TES electron system, generating a fast ``athermal'' signal that happens before the usual thermal one (related to the absorber temperature rise). Our detectors are designed to optimize the collection of this fast signal in order to speed up the response time, an important feature for an anticoincidence detector. In this respect, the AC-S7 TES network was designed to ensure a large and uniform surface coverage for an efficient athermal phonons collection,  while constraining down the heat capacity contribution of the metal film due to the quite small TES size. 

To investigate the possibility to further improve the athermals collecting efficiency, the AC-S8 sample is featured also with a network of Al-fingers directly connected to the TES. The athermal phonons can be absorbed in the Al film, breaking Cooper pairs and generating electron-like quasiparticles. The quasiparticles diffuse through the fingers towards the TES, where they deposit their energy contributing to the fast athermal signal\cite{al}. 

\begin{figure}[htbp]
\begin{center}
\includegraphics[width=0.8\linewidth]{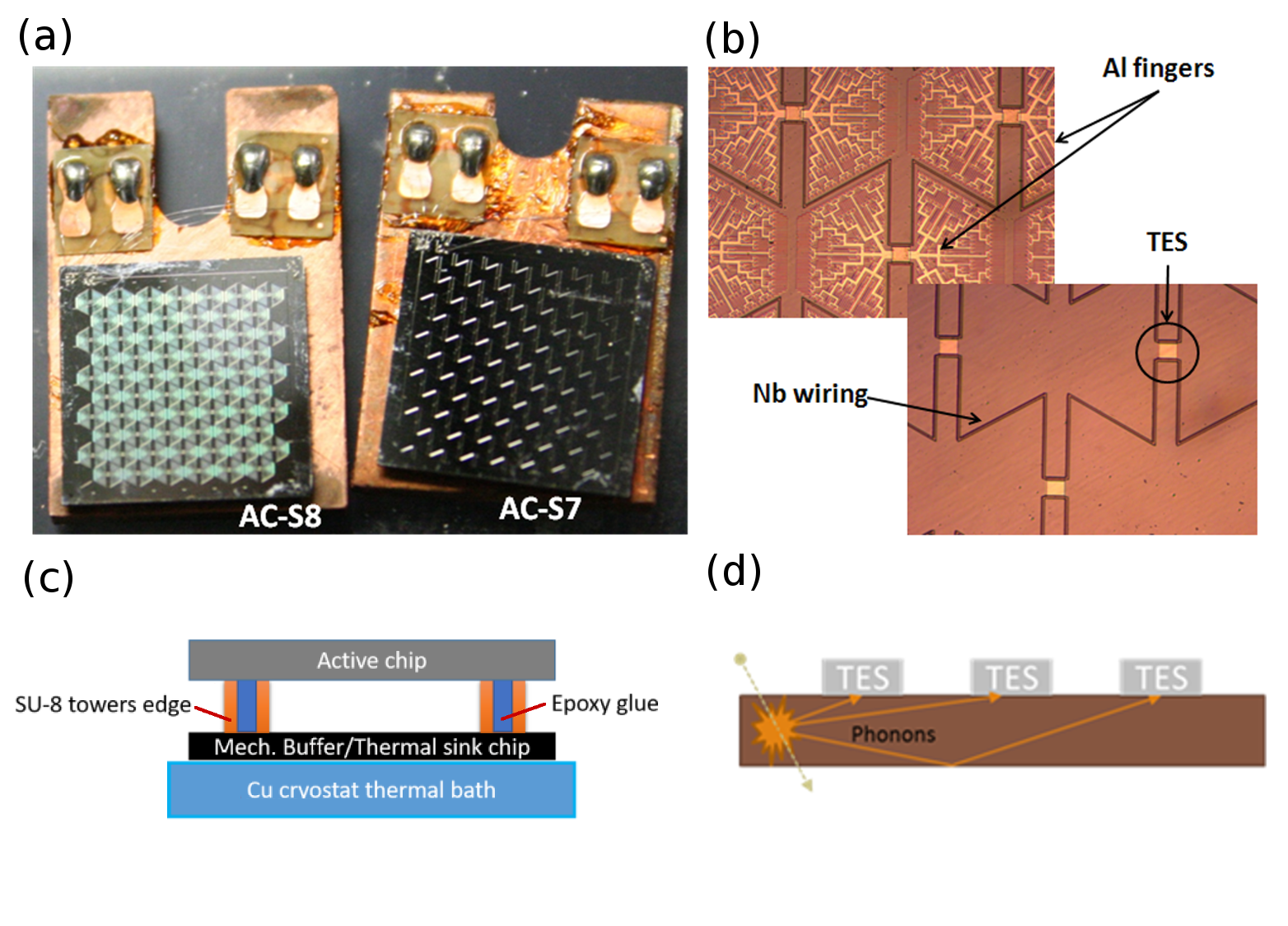}
\caption{\textit{(a):} The AC-S8 and AC-S7 samples. 
\textit{(b):} Details about the TES network, Nb wiring and Al fingers. 
\textit{(c):} Scheme of ''tower'' geometry with highlighted absorber, epoxy thermal link and buffer chip.
\textit{(d):} Scheme of the detection principle: the first burst of ballistic phonons could be directly collected by the TES network.
(Color figure online)}
\end{center}
\label{fig:ACS7and8}
\end{figure}

\begin{table}[htbp]
\centering
\caption{Main parameters of the AC-S7 and AC-S8 samples.}
\label{tab:ACS7-8}
\begin{tabular}{ll}
\hline\noalign{\smallskip}
Parameter & Value \\
\noalign{\smallskip}\hline\noalign{\smallskip}
Absorber Silicon size &  10x10 mm$^2$, 380 $\mu$m thick \\
TES (x65) Iridium size &  100x100 $\mu$m$^2$, $\sim $200 nm thick \\
Niobium wiring & $\sim$ 870 nm thick\\ 
Aluminum fingers (only on AC-S8) &  $\sim$ 420 nm thick \\
\noalign{\smallskip}\hline
\end{tabular}
\end{table}

\noindent In the next sections we report the measurements performed on the AC-S8 prototype in order to preliminary characterize the sample and investigate the role of the additional Aluminum finger network.

\section{AC-S8 transition and I-V characteristic curves}

The AC-S8 prototype has been tested in the INAF/IAPS Roma CryoLab using a dry cryogenic system (an Adiabatic Demagnetization Refrigerator precooled with a Pulse Tube cooler). The sample was magnetically shielded by a 1 mm thick Cryophy\cite{cryophy} shield anchored to the 2.5 K stage of the cryostat. The detector has been coupled to a commercial SQUID array chip including a $R_S=$0.2 m$\Omega$ shunt resistor (Magnicon series array C6X16F), and operated with a Magnicon XXF-1 electronics.

\begin{figure}[htbp]
\begin{center}
\includegraphics[width=0.43\linewidth]{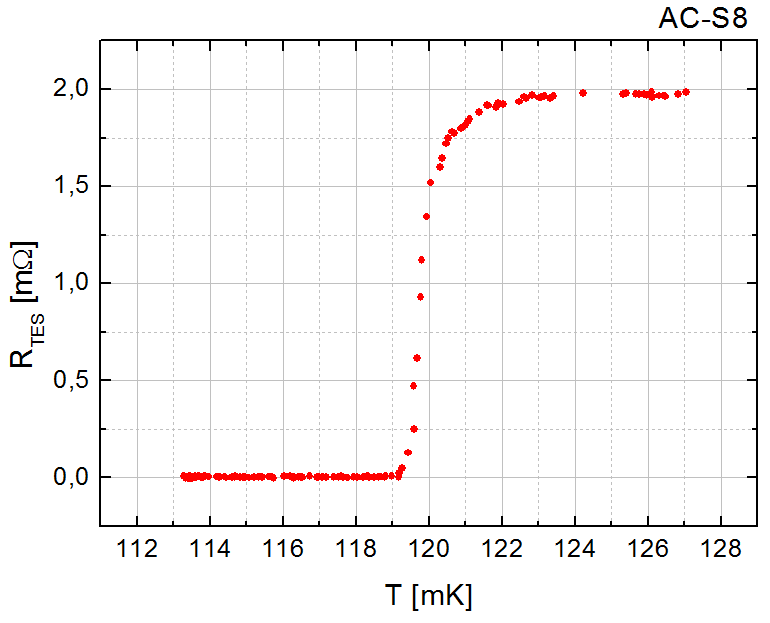}
\includegraphics[width=0.56\linewidth]{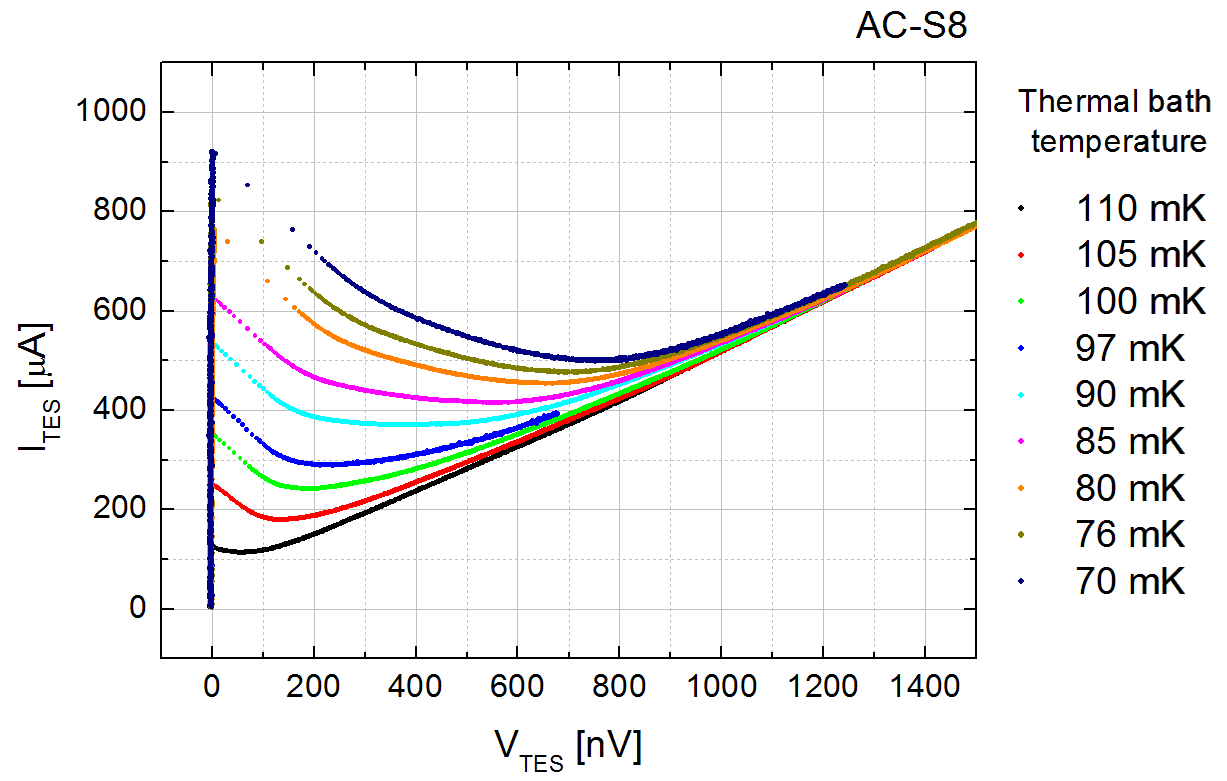}
\caption{Characterization measurements performed on the AC-S8 sample. \textit{Left:} Transition curve.
\textit{Right:} I-V curves. (Color figure online)}
\end{center}
\label{RTIV}
\end{figure}

The transition curve of the sample is shown in Fig.~2-Left. The measurement has been performed biasing the sensor with a low frequency sine wave (f = 22Hz)  and sweeping the bath temperature. The bias current was kept small (few $\mu A_{PP}$) to minimize the TES self-heating and avoid systematic effects. The plot shows a normal resistence R$_{N}$ = 2.0 m$\Omega$, a transition temperature T$_{C}$ = 120 mK (evaluated at R$_{TES}$/R$_{N}$ = 0.5) and a 10\%-90\% transition width $\Delta T_{C, 10\%-90\%} <$ 2 mK. The narrowness of the transition proves the homogeneity  of the TES network, which we remark is constituted by 65 identical Ir films connected in parallel, validating the sample manufacturing process. 

In Fig.~2-Right are reported some characteristics I-V curves of the prototype. The curves have been acquired for decreasing voltage values at fixed thermal bath temperatures. The wide explored temperature range (from T$_{B}$ = 70 mK to T$_{B}$ = 110 mK) shows that the detector can be operated with the thermal bath quite far from the transition, until a maximum temperature difference $\Delta T_{max} = (T_{C} - T_{B})_{max} \sim$ 50 mK. This is an important step in the path towards the DM development, since it will operate with a similar $\Delta T$: for the CryoAC DM are indeed planned T$_{B}$ = 50-55 mK and T$_{C}$ = 90 mK TBC (Ir/Au bilayer TES).

\section{The pulse dynamic}

The AC-S8 detector has been illuminated with a low activity $^{241}$Am gamma source. The source was shielded with a copper sheet (0.5 mm thick) in order to provide only 60 keV photons, with an expected count rate of $\sim$ 3 cps. The bath temperature was set to 70 mK and the detector biased at R$_{TES}$/R$_N$ = 0.3 (V$_{TES}$ = 340 nV, I$_{TES}$ = 615 $\mu A$).
We remark that our detectors, due to the high number of TES, require high current (mA) to overcome the critical current and drive the TES from superconductive to the transition region. Thus, with respect to our previous test procedures, we have driven the TES to the normal state by increasing the bath temperature, injected the bias current, then decreased the thermal bath down to a good working point.
During 400 s of acquisition, 1244 pulses have been triggered, so having a count rate consistent with the expected one. The average 60 keV pulse is shown in Fig.~3.  The result of a double pulse fitting procedure is overplotted, disentangling the expected fast athermal and the slow thermal components of the signal. The fit function is: 

\begin{equation}
I(t) = PH_{ath} \cdot ( e^{-t/\tau_{D, ath}} - e^{-t/\tau_{R, ath}} ) + PH_{th} \cdot ( e^{-t/\tau_{D, th}} - e^{-t/\tau_{R, th}} )
\end{equation}
where for both the pulse components PH, $\tau_R$ and $\tau_D$ represent the Pulse Height and the characteristic rise and decay times. 

\begin{figure}[htbp]
\begin{center}
\includegraphics[width=0.60\linewidth]{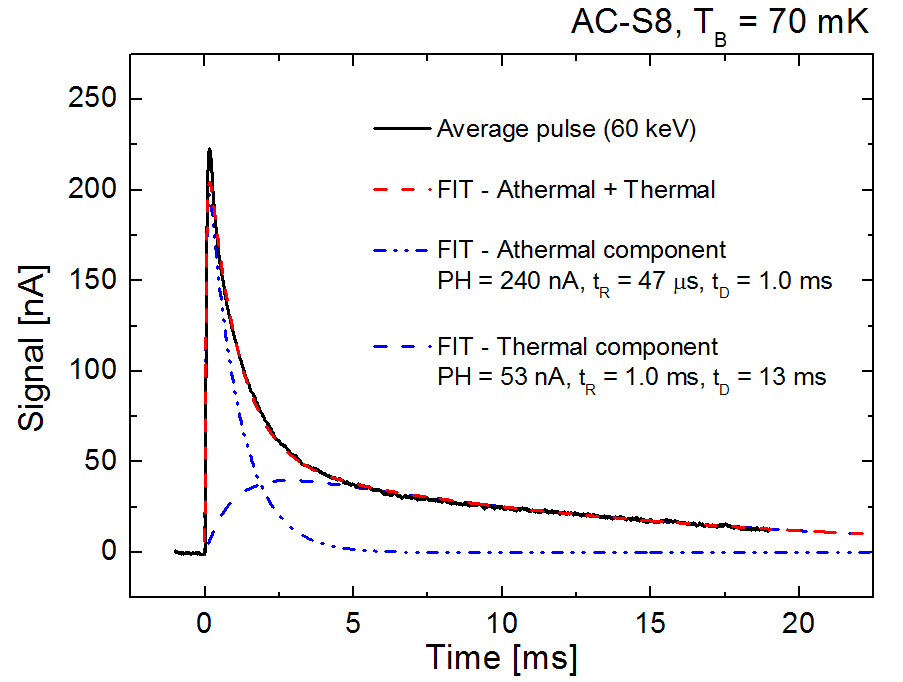}
\caption{Average 60 keV pulse acquired illuminating AC-S8 with a $^{241}$Am source. The result of the double pulse fitting procedure is overplotted. (Color figure online)}
\label{avpulse}
\end{center}
\end{figure}

\subsection{From AC-S3 to AC-S8: the pulse rising}

The measured average pulse athermal rise time $\tau_{R,ath}$ = 47 $\mu$s (Fig.~3) is compatible with the L/R $\sim$ 50 $\mu$s  of the TES circuit. This was evaluated taking into account the L = 90 nH total inductance and the $R_{P}$ = 1.0 m$\Omega$ parasitic resistance measured in series with the TES. The fast athermal rise time shows that the configuration of Al fingers does not slow down the rising of the pulse, as was instead observed in different prototypes developed in the past. Consider for comparison AC-S3, an old sample with 3 rectangular Al islands (2x5 mm$^2$) connecting 4 Ir/Au TES (each one 1x1.5 mm$^2$). In that case the wide area athermals collectors permitted the quasiparticle recombination during their diffusion in the aluminum, generating a late family of phonons\cite{acs3}. This phenomenon is well highlighted in Fig.~4, where the scatter plots \textit{Pulse integral} vs \textit{Rise time} of the 60 keV pulses acquired by AC-S3 and AC-S8 are shown. In the old sample two different families of pulses are evident (Fig.~4-Left). The slowest family is due to the phonons generated by the quasiparticle recombination in the Al collectors. The AC-S8 design prevents this issue, being based on a tree-like network of shorter Al fingers (max size 25x500 $\mu$m$^2$). In this way the quasiparticle path towards the TES is minimized, and the probability of recombination significantly reduced. As a result, the slow family of pulses is not present (Fig.~4-Right).\newline

\begin{figure}[htbp]
\begin{center}
\includegraphics[width=0.99\linewidth]{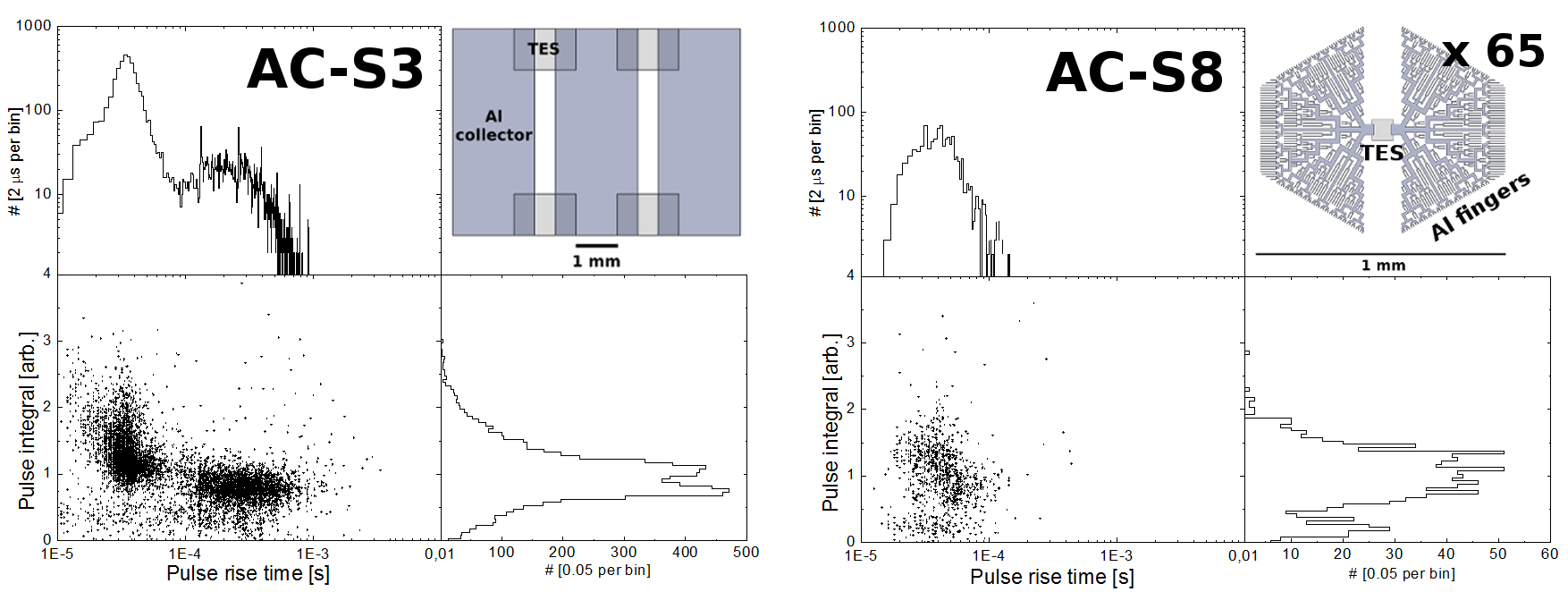}
\caption{\textit{Left:} Scatter plot of the ''Pulse integral'' vs ''Rise time'' of the 60 keV pulses acquired by ACS3. Two different families of pulses are evident. \textit{Right:} The same plot referring to the pulses acquired by AC-S8. In this case the slow family of pulses is not present. (Color figure online)}
\end{center}
\label{scatters}
\end{figure}

The distribution of the in-time integral of the pulses in Fig.~4-Right represents the preliminary energy spectrum acquired by AC-S8, with the main peak corresponding to the 60 keV events. The spectrum shows a very poor energy resolution ($\Delta$E/E $\sim$ 0.8 at 60 keV), making difficult to provide an assessment of the detector low energy threshold. This is due to the fact that the detector has been not operated in good conditions. First, the high value of the parasitic resistance ($R_P$ = 1 m$\Omega$ = 0.5$R_N$) strongly affected the electrothermal feedback gain loop. This resistance was located between the input coil and the TES, at the contact between our superconducting wiring and the Nb pads of the SQUID packaging. Further, the shunt resistance on board the SQUID chip (at 500 mK) generated a high Johonson noise.

The foreseen DM setup will solve these issues, having the SQUID close to the detector (few mm, at the 50 mK stage), and the connection between the input coil and the TES realized with ultrasonic aluminum wire bonding. Furthermore, the DM normal resistance is expected to be higher than the AC-S8 one ($R_{N, DM} \sim$ 10 m$\Omega$), thanks to a different TES aspect ratio.

We remark that the setup limitations do not affect the evaluation of the AC-S8 athermals collection efficiency reported in the next section, which has been performed by means of an average pulse analysis and it was found to be independent from the detector working point.

\subsection{From AC-S7 to AC-S8: the improvement in the athermals collection}
From the double-component fit in Fig.~3, it is possible to evaluate the ratio between the athermal and the total energy associated to the average pulse, which are proportional to the integral of the respective fit functions.
We found $\varepsilon_{AC-S8}$ = $E_{ath}/(E_{ath} + E_{th}) |_{AC-S8}$ =~26\%, more than a factor 4 above the AC-S7 value $\varepsilon_{AC-S7} \sim 6\%$\cite{PCA}. We interpret this as an evidence of the higher AC-S8 athermals collection efficiency due to the Al finger network. 

To consolidate this result we have repeated the average pulse analysis biasing the detector at different transition levels and with different thermal bath temperatures. The results are reported in Fig.~5, which shows that the $\varepsilon$ ratio is roughly independent from the detector working point, confirming the efficiency increasing.

\begin{figure}[htbp]
\begin{center}
\includegraphics[width=0.65\linewidth]{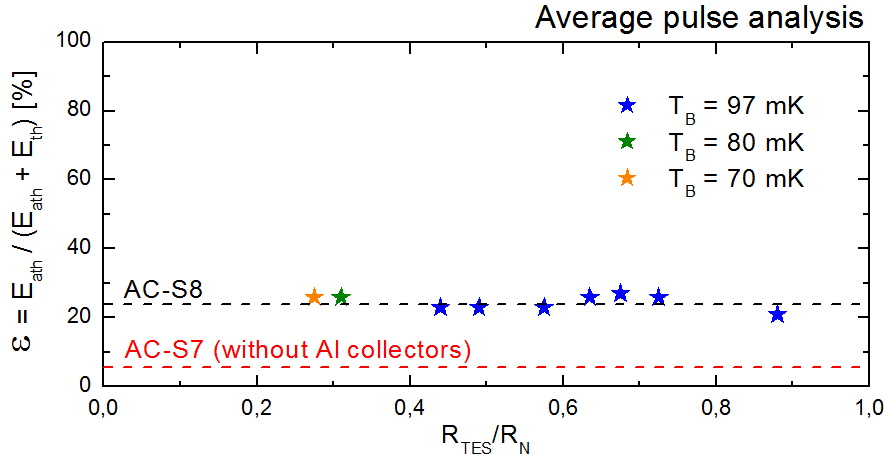}
\caption{Ratio between the athermal and the total (athermal + thermal) energy of the average 60 keV pulses acquired with AC-S8 in different working points. The mean value is compared with the AC-S7 one (from\cite{PCA}). (Color figure online)}
\end{center}
\label{eps}
\end{figure}

\section{Conclusions}

We have here reported the main results of the test activities performed on AC-S8, the last single pixel prototype in the path towards the CryoAC DM development, that has on board aluminum fingers to increase the athermal collection efficiency with respect to the AC-S7 twin detector\cite{LTD16}.

We have been able to operate the detector with the thermal bath quite far from the transition (until $\sim$ 50 mK below), as requested for the DM. This is the first time that we have been able to operate a wide absorber detector, having on-board ten's of TES, at a thermal bath temperature quite far from the critical temperature (Tc $\sim$ 120 mK). However, we underline that to deal with the high current needed to bias the detector ($I_{TES}$= 615 $\mu$A for AC-S8, corresponding to $I_{B}$= 5.5 mA to the whole TES bias circuit), in the current DM design it is foreseen an additional platinum heater deposited on the absorber. In this way, it will be possible to operate the detector also  by smaller bias current, locally increasing the absorber temperature with the on-board heater. The heater can also be used just to drive the TES into the normal state and bias it without overcoming the high critical current (some mA), and then turned off during the detector operations. 

As expected, the AC-S8 pulse dynamic is characterized by the presence of two different components: the fast athermal component (due to the collection of high energy ballistic phonons) and the slow thermal one. We have shown that the detector design is able to prevent the quasi-particles recombination in the phonon collectors, assuring a fast pulse rising front (L/R limited), thanks to a careful design based on quite small finger size. Furthermore, we have obtained as expected a first evidence of the improvement in the athermal collection efficiency due to the use of the aluminum fingers network (from AC-S7 $\sim$ 6$\%$ to AC-S8 up to $\sim$ 26$\%$).

These results from the AC-S8 detector, merged with the ones from AC-S7\cite{LTD16}, 
will help in generating the CryoAC instrument requirements.

We also report that, in order to reach a better understanding of the detector physics, we have started an activity aimed at developing a model of the detector to properly describe the pulse athermal/thermal dynamic, as a due activity in the context of the usual Phase A mission programm (trade-off study). This is fundamental with respect to the evaluation of the detector rejection efficiency, which is one of the key parameters of the CryoAC.

\begin{acknowledgements}
This work has been partially supported by ASI (Italian Space Agency) through the Contract no. 2015-046-R.0, and by ESA (European Space Agency) through the Contract no. 4000114932/15/NL/BW.
\end{acknowledgements}

\pagebreak

\end{document}